\documentclass[nofootinbib,prd,twocolumn,showpacs,showkeys,preprintnumbers]{revtex4}
\usepackage{hyperref,amssymb,amsmath,mathrsfs,bm,graphicx}
\begin{document}
\title {Complexity hierarchies in Euclidean stars}
\author{L. Herrera}
\email{lherrera@usal.es}
\affiliation{Instituto Universitario de F\'isica
Fundamental y Matem\'aticas, Universidad de Salamanca, Salamanca 37007, Spain}
\author{A. Di Prisco}
\email{alicia.diprisco@ucv.ve}
\affiliation{Escuela de F\'\i sica, Facultad de Ciencias, Universidad Central de Venezuela, Caracas 1050, Venezuela}
\author{J. Ospino}
\email{j.ospino@usal.es}
\affiliation{Departamento de Matem\'atica Aplicada and Instituto Universitario de F\'isica
Fundamental y Matem\'aticas, Universidad de Salamanca, Salamanca 37007, Spain}
\date{\today}
\begin{abstract}
 We establish a hierarchy of  Euclidean stars according to their degree of complexity, as measured by the complexity factor and the complexity of the pattern of evolution. We consider  both, non--dissipative and dissipative systems. Solutions are ranged from the simplest one, in order of increasing complexity. Some specific models are found and analyzed in detail.
\end{abstract}
\maketitle
\section{Introduction}
This work deals with  a family of solutions to Einstein equations for fluid distributions  satisfying  the heuristic assumption that the areal radius of  any  shell of fluid, which is the radius obtained from its area, equals the proper radial distance from the center to the shell \cite{pe}, thereby named ``Euclidean stars'' (see \cite{5e,9e,6e,7e,8e,4e,2e,3e,1e}  for generalizations and/or applications of this idea).

Since the  above mentioned assumption is obviously satisfied in the weak field (Newtonian) limit, we expect that Euclidean stars could provide simple models of evolving stars, 
 which are relatively simple to analyze but still contain some of the
essential features of a realistic situation.

In this work  we endeavor to classify  all possible solutions describing ``Euclidean stars'',   according  to the complexity of their structure and their pattern of evolution. The simplest solution satisfies  the condition of vanishing complexity factor and evolves in the homologous or quasi-homologous regime. From there on we  obtain different families of solutions by relaxing the above conditions. 

A rigorous definition of complexity is important  since complexity,  however we define  it, is a physical concept   deeply intertwined with fundamental aspects of   the system, which we expect that could provide relevant information  about its behavior.

Here we shall resort to the  definition given in \cite{6n},  intuitively associated to the very concept of ``structure'' within the fluid distribution. Such a definition is based on the  assumption that the simplest system (or at least one of them) is represented by the homogeneous fluid with isotropic pressure. Having assumed this conjecture for  a vanishing complexity system, the variable responsible for measuring complexity, which we call the complexity factor, appears in a natural way in the orthogonal splitting of the Riemann tensor  (see \cite{6n} for details).

In the case of  time-dependent  systems, there is another aspect that requires our attention regarding the degree of complexity of the system. We have in mind the problem of the complexity of the patterns of evolution of the fluid.  

In \cite{7n}  it was shown that the homologous condition ($H$)  may be suitable to  characterize the simplest possible mode of evolution. It is worth recalling that in Newtonian physics this condition implies that the radial velocity of any fluid element is proportional to its radial distance from the center, implying in its turn that the ratio of the radii of two comoving shells of fluid remains constant all along the evolution. However, in the relativistic regime these two properties of homologous condition are independent.  Thus, we may consider a somewhat less restrictive condition by assuming only the first of the two above mentioned features.This is   the so-called quasi-homologous condition ($QH$)  \cite{7nn} (see details in Section 3.2).

Thus we shall select solutions corresponding to Euclidean stars, and range them from the simplest one satisfying the vanishing complexity factor condition and evolving in the homologous regime, to solutions with an increasing level of complexity. 

We shall consider dissipative and non-dissipative models.

The inclusion of dissipation is justified by the fact that gravitational collapse is a
highly dissipative process (see \cite{Mitra} and references
therein). This dissipation is required to account for the very large
(negative) binding energy of the resulting compact object of the order
of $-10^{53}$ erg \cite{nr1}. 

Non-dissipative Euclidean models are necessarily geodesic, belonging to the Lema\^{\i}tre-Tolman-Bondi (LTB) solutions \cite{lemaitre,tolman,bondi} (more specifically to the parabolic subclass). They may describe evolving  dust or, more generally,  evolving  anisotropic fluids \cite{sussman}.

We shall first consider the non--dissipative case, and shall identify the simplest model according to the criteria mentioned above. Such solution appears to be the well known Friedman-Robertson-Walker space-time (FRW). Next, we shall relax the $H$ or the $QH$ conditions, keeping only the vanishing complexity factor condition. This leads to a family of solutions depending on two arbitrary functions of the radial coordinate. A specific model of this family is found and analyzed in detail.

Next we shall consider the dissipative case. We shall first notice that we cannot impose the $H$ condition since it would imply that the fluid is geodesic, which in the context of Euclidean stars implies that it is non-dissipative. Thus, we shall first  consider the case when the vanishing complexity factor condition and $QH$ hold, which leads to a large family  of solutions, a  specific example of which will be analyzed in detail.

Finally we shall relax the $QH$ condition,  keeping only  the vanishing complexity factor condition. Additionally, in order to be able to integrate the system,  we  shall assume the fluid to be shear-free. These two conditions lead to a large family of solutions already known in the literature \cite{hlsw, hja, GM}.

\section{The Euclidean condition and its consequences}
We shall start by defining all concepts, variables and equations required for our discussion.This includes the definitions  of complexity, the matching conditions, the Weyl tensor and the transport equation. Next, we shall define the Euclidean condition, and the general properties of fluids satisfying this condition. 
\subsection{Basic equations and variables}
Let us  consider  spherically symmetric distributions  of 
fluid, bounded by a spherical surface $\Sigma$. The fluid is
assumed to be locally anisotropic with principal stresses unequal and undergoing dissipation in the
form of heat flow.
Choosing comoving coordinates inside $\Sigma$, the general
interior metric can be written as
\begin{equation}
ds^2=-A^2dt^2+B^2dr^2+R^2(d\theta^2+\sin^2\theta d\phi^2),
\label{1}
\end{equation}
where $A$, $B$ and $R$ are functions of $t$ and $r$ and are assumed
positive. We number the coordinates $x^0=t$, $x^1=r$, $x^2=\theta$
and $x^3=\phi$.

The matter energy-momentum tensor $T_{\alpha\beta}$ inside $\Sigma$
has the form
\begin{equation}
T_{\alpha\beta}=(\mu +
P_{\perp})V_{\alpha}V_{\beta}+P_{\perp}g_{\alpha\beta}+(P_r-P_{\perp})\chi_{
\alpha}\chi_{\beta}+q_{\alpha}V_{\beta}+V_{\alpha}q_{\beta}, \label{3}
\end{equation}
where $\mu$ is the energy density, $P_r$ the radial pressure,
$P_{\perp}$ the tangential pressure, $q^{\alpha}$ the heat flux,
$V^{\alpha}$ the four-velocity of the fluid and
$\chi^{\alpha}$ a unit four-vector along the radial direction.

It will be convenient to express the  energy momentum tensor  (\ref{3})  in the equivalent (canonical) form
\begin{equation}
T_{\alpha \beta} = {\mu} V_\alpha V_\beta + P h_{\alpha \beta} + \Pi_{\alpha \beta} +
q \left(V_\alpha \chi_\beta + \chi_\alpha V_\beta\right), \label{Tab}
\end{equation}
with
$$ P=\frac{P_{r}+2P_{\bot}}{3}, \qquad h_{\alpha \beta}=g_{\alpha \beta}+V_\alpha V_\beta,$$

$$\Pi_{\alpha \beta}=\Pi\left(\chi_\alpha \chi_\beta - \frac{1}{3} h_{\alpha \beta}\right), \qquad \Pi=P_{r}-P_{\bot}.$$
These quantities satisfy
\begin{equation}
V^{\alpha}V_{\alpha}=-1, \;\; V^{\alpha}q_{\alpha}=0, \;\; \chi^{\alpha}\chi_{\alpha}=1, \;\;
\chi^{\alpha}V_{\alpha}=0.
\end{equation}
Since we assumed the metric (\ref{1}) comoving then
\begin{equation}
V^{\alpha}=A^{-1}\delta_0^{\alpha}, \;\;
q^{\alpha}=qB^{-1}\delta^{\alpha}_1, \;\;
\chi^{\alpha}=B^{-1}\delta^{\alpha}_1, \label{5}
\end{equation}
where $q$ is a function of $t$ and $r$.

The four-acceleration $a_{\alpha}$ and the expansion $\Theta$ of the fluid are
given by
\begin{equation}
a_{\alpha}=V_{\alpha ;\beta}V^{\beta}, \;\;
\Theta={V^{\alpha}}_{;\alpha}, \label{4b}
\end{equation}
and its  shear $\sigma_{\alpha\beta}$ by
\begin{equation}
\sigma_{\alpha\beta}=V_{(\alpha
;\beta)}+a_{(\alpha}V_{\beta)}-\frac{1}{3}\Theta(g_{\alpha\beta}+V_{\alpha}V
_{\beta}),
\label{4a}
\end{equation}
where  semicolon denotes covariant derivative, and the round bracket around the indexes denotes symmetrization.

 From (\ref{4b}) with (\ref{5}) we have for the  four-acceleration and its scalar $a$,
\begin{equation}
a_1=\frac{A^{\prime}}{A}, \;\; a^2=a^{\alpha}a_{\alpha}=\left(\frac{A^{\prime}}{AB}\right)^2, \label{5c}
\end{equation}
and for the expansion
\begin{equation}
\Theta=\frac{1}{A}\left(\frac{\dot{B}}{B}+2\frac{\dot{R}}{R}\right),
\label{5c1}
\end{equation}
where the  prime stands for $r$
differentiation and the dot stands for differentiation with respect to $t$.
Using  (\ref{5}) and  (\ref{4a}) we may write
\begin{equation}
\sigma_{\alpha \beta}=\sigma\left(\chi_\alpha \chi_\beta-\frac{h_{\alpha \beta}}{3}\right),
\label{sigma}
\end{equation}
where
\begin{equation}
\sigma=\frac{1}{A}\left(\frac{\dot{B}}{B}-\frac{\dot{R}}{R}\right).\label{5b1}
\end{equation}
The mass function $m(t,r)$ introduced by Misner and Sharp
\cite{13n} (see also \cite{13nn}) reads
\begin{equation}
m=\frac{R^3}{2}{R_{23}}^{23}
=\frac{R}{2}\left[\left(\frac{\dot R}{A}\right)^2-\left(\frac{R^{\prime}}{B}\right)^2+1\right].
 \label{17masa}
\end{equation}

 As it follows at once from (\ref{1}), the function $R$ is the areal radius (i.e. the radius as measured from the area of a spherical surface).  Therefore we  can define the velocity $U$ of the evolving
fluid  as the variation of the areal radius with respect to proper time, i.e.
\begin{equation}
U=D_TR, \label{19n}
\end{equation}
where $D_T=(1/A)(\partial/\partial t)$ defines the derivative with respect to proper time.
Then (\ref{17masa}) can be rewritten as
\begin{equation}
E \equiv \frac{R^{\prime}}{B}=\left(1+U^2-\frac{2m}{R}\right)^{1/2}.
\label{20x}
\end{equation}

\subsection{The Weyl tensor}
Let us recall that in  the spherically symmetric case the Weyl tensor  ($C^{\rho}_{\alpha
\beta
\mu}$) is   defined by its ``electric'' part  $E_{\gamma \nu }$ alone, since its  ``magnetic'' part
vanishes.
Indeed, the magnetic part of the Weyl tensor is associated with vorticity, which of course is absent in any spherically symmetric distribution.

Thus we have
  \begin{equation}
E_{\alpha \beta} = C_{\alpha \mu \beta \nu} V^\mu V^\nu,
\label{elec}
\end{equation}
whose non trivial components are
\begin{eqnarray}
E_{11}&=&\frac{2}{3}B^2 {\cal E},\nonumber \\
E_{22}&=&-\frac{1}{3} R^2 {\cal E}, \nonumber \\
E_{33}&=& E_{22} \sin^2{\theta},
\label{ecomp}
\end{eqnarray}
where

\begin{eqnarray}
{\cal E}&=& \frac{1}{2 A^2}\left[\frac{\ddot R}{R} - \frac{\ddot B}{B} - \left(\frac{\dot R}{R} - \frac{\dot B}{B}\right)\left(\frac{\dot A}{A} + \frac{\dot R}{R}\right)\right]\nonumber \\&+& \frac{1}{2 B^2} \left[\frac{A^{\prime\prime}}{A} - \frac{R^{\prime\prime}}{R} + \left(\frac{B^{\prime}}{B} + \frac{R^{\prime}}{R}\right)\left(\frac{R^{\prime}}{R}-\frac{A^{\prime}}{A}\right)\right]\nonumber \\&- &\frac{1}{2 R^2}.
\label{E}
\end{eqnarray}
Observe that  the electric part of the 
Weyl tensor may also be written as
\begin{equation}
E_{\alpha \beta}={\cal E} (\chi_\alpha \chi_\beta-\frac{1}{3}h_{\alpha \beta}).
\label{52}
\end{equation}
\subsection{The transport equation}
\label{sec:4}
 In the diffusion approximation we shall need a transport equation to evaluate the temperature and its evolution within the fluid distribution. Here we shall resort to a transport equation derived from a causal  dissipative theory ( e.g. the
Israel-Stewart second
order phenomenological theory for dissipative fluids \cite{14n,15n,16n}).

Thus the  corresponding  transport equation for the heat flux reads
\begin{equation}
\tau
h^{\alpha\beta}V^{\gamma}q_{\beta;\gamma}+q^{\alpha}=-\kappa h^{\alpha\beta}
(T_{,\beta}+Ta_{\beta}) -\frac 12\kappa T^2\left( \frac{\tau
V^\beta }{\kappa T^2}\right) _{;\beta }q^\alpha ,  \label{21t}
\end{equation}
where $\kappa $  denotes the thermal conductivity, and  $T$ and
$\tau$ denote temperature and relaxation time respectively. Observe
that, due to the symmetry of the problem, equation (\ref{21t}) only
has one independent component, which may be written as
\begin{equation}
\tau{\dot q}=-\frac{1}{2}\kappa qT^2\left(\frac{\tau}{\kappa
T^2}\right)^{\dot{}}-\frac{1}{2}\tau q\Theta A-\frac{\kappa}{B}(TA)^{\prime}-qA.
\label{te}
\end{equation}
In the case $\tau=0$ we recover the Eckart--Landau equation \cite{17T, 11L}.

 For simplicity one can consider in some cases  the so called ``truncated'' version where the last term in (\ref{21t}) is neglected \cite{17n},
\begin{equation}
\tau
h^{\alpha\beta}V^{\gamma}q_{\beta;\gamma}+q^{\alpha}=-\kappa h^{\alpha\beta}
(T_{,\beta}+Ta_{\beta}) \label{V1},
\end{equation}
and whose    only non--vanishing independent component becomes
\begin{equation}
\tau \dot q+qA=-\frac{\kappa}{B}(TA)^{\prime}. \label{V2}
\end{equation}

 \subsection{The exterior spacetime and junction conditions}
Since our fluid distribution is bounded we assume that  outside $\Sigma$ the space--time is described  by  Vaidya metric which reads.
\begin{equation}
ds^2=-\left[1-\frac{2M(v)}{\bf r}\right]dv^2-2d{\bf r}dv+{\bf r}^2(d\theta^2
+\sin^2\theta
d\phi^2) \label{1int},
\end{equation}
where $M(v)$  denotes the total mass, ${\bf r}$ is a null coordinate
and  $v$ is the retarded time.

The smooth matching of the full nonadiabatic sphere   to
the Vaidya spacetime, on the surface $r=r_{\Sigma}=$ constant, requires the fulfillment of the Darmois conditions \cite{14nn}, i.e. the continuity of the first and second fundamental forms across $\Sigma$ (see \cite{chan} and references therein for details), which implies 
 \begin{equation}
m(t,r)\stackrel{\Sigma}{=}M(v), \label{junction1}
\end{equation}
and

\begin{equation}
q\stackrel{\Sigma}{=}P_r,\label{j3}
\end{equation}
where $\stackrel{\Sigma}{=}$ means that both sides of the equation
are evaluated on $\Sigma$.

When  Darmois conditions are not satisfied  the  boundary surface is a thin shell \cite{15nn}.
\section{The conditions for classifying Euclidean stars}
As mentioned before we endeavor to classify Euclidean stars  according to their degree of complexity.  Such a complexity will be measured by the complexity factor, and the complexity of the pattern of evolution. We shall next provide a brief description of the complexity factor,  the homologous and the quasi-homologous regime which are the two modes of evolution assumed to be the simplest ones.

\subsection{The   complexity factor}
The complexity factor is  a scalar function that has been proposed in order  to measure the degree of complexity of a given fluid distribution \cite{6n, 7n}.

The complexity factor is identified with the scalar function $Y_{TF}$ which defines the trace--free part of the electric Riemann tensor (see \cite{7n} for details).

It can be expressed in terms of physical variables as 

\begin{equation}
Y_{TF}= - 8 \pi \Pi   + \frac{4 \pi}{R^3}\int^r_0{R^3\left(\mu^\prime - \frac{3  q BU}{R}\right) dr},
\label{Yi}
\end{equation}
or in  terms of the metric functions

\begin{eqnarray}
Y_{TF}= \frac{1}{A^2}\left[\frac{\ddot R}{R} - \frac{\ddot B}{B} + \frac{\dot A}{A}\left(\frac{\dot B}{B} - \frac{\dot R}{R}\right)\right]\nonumber \\+ \frac{1}{ B^2} \left[\frac{A^{\prime\prime}}{A} -\frac{A^{\prime}}{A}\left(\frac{B^{\prime}}{B}+\frac{R^{\prime}}{R}\right)\right],
\label{itfm}
\end{eqnarray}
or    
 \begin{equation}
Y_{TF}={\cal E}-4\pi \Pi .\label{EY}
\end{equation}

\subsection{The homologous ($H$) and the quasi-homologous ($QH$) condition}
\label{sec:3}

The $QH$ condition is a generalization of the $H$ condition.  This latter condition has been assumed in \cite{7n} to represent  the simplest mode of evolution of the fluid distribution. However it appears  to be too   stringent thereby  excluding many potential interesting scenarios. Therefore in \cite{7nn} we have proposed to relax ($H$), and consider what we  called  the  ``quasi--homologous'' condition ($QH$), which corresponds to a more complex pattern of evolution than $H$. 

More specifically, the $H$ condition implies that 
\begin{equation}
U=\omega (t) R, \qquad \omega (t)\equiv\frac{U_{\Sigma}}{R_\Sigma},
\label{h1n}
\end{equation}
and 
 
 \begin{equation}
\frac{R_I}{R_{II}}=\mbox{constant},
\label{hn2}
\end{equation}
where $R_I$ and $R_{II}$ denote the areal radii of two concentric shells ($I,II$) described by $r=r_I={\rm constant}$, and $r=r_{II}={\rm constant}$, respectively. 
 
 These  relationships  are characteristic of the homologous evolution in Newtonian hydrodynamics \cite{22n,20n,21n}. More so, in this latter case (\ref{h1n}) implies (\ref{hn2}). However in the relativistic case both (\ref{h1n}) and  (\ref{hn2}) are in general independent, and the former implies the latter only in very special cases (e.g. the geodesic case).
 
On the other hand $QH$ only requires (\ref{h1n}), which using the field equations may  also be written as (see \cite{7nn} for details)
\begin{equation}
\frac{4\pi}{R^\prime}B  q+\frac{\sigma}{ R}=0.
\label{ch1}
\end{equation}
\subsection{Euclidean stars and their properties}
Regularity condition at the center requires that the perimeter of an infinitesimally small circle around the center be given by $2 \pi l$
where $l$ is the proper radius of the circle, i.e. the proper radial distance from the center to the circle, given  by
 \begin{equation}
l=\int_0^r Bdr.
\label{ml}
\end{equation}
On the other hand, the perimeter of any circle, as it follows from (\ref{1}) is given by $ 2 \pi R$
implying that in the neighborhood of the center regularity conditions imply 
\begin{equation}
R^{\prime\stackrel{r=0}{=}}B.
\label{nc0}
\end{equation}

Euclidean stars are characterized by the fulfillment of the above condition  not only at the center of symmetry, but   at  all points of the fluid distribution.

Thus we shall assume that the ``areal'' radius ($R$) representing the radius as measured by its spherical surface, and the {\it proper radius}  defined by  $\int B(t,r)dr$,  be equal. These two radii in general, in Einstein's theory, need not to be equal, unlike in Newton's theory. Such a condition defines the Euclidean star.
 Hence with this condition we can write,
\begin{equation}
B=R^{\prime}, \label{16}
\end{equation}
implying from (\ref{20x})
\begin {equation}
E=1.\label{21E}
\end{equation}

The field equations with this condition  become
\begin{eqnarray}
8\pi\mu =\frac{1}{A^2}\left(\frac{\dot R}{R}+2\frac{{\dot R}^{\prime}}{R^{\prime}}\right)\frac{\dot R}{R}, \label{17}\\
8\pi qAR^{\prime}=-2\frac{\dot R}{R}\frac{A^{\prime}}{A}, \label{18}\\
8\pi P_r=-\frac{1}{A^2}\left[2\frac{\ddot R}{R}-\left(2\frac{\dot A}{A}-\frac{\dot R}{R}\right)
\frac{\dot R}{R}\right]+2\frac{A^{\prime}}{A}\frac{1}{RR^{\prime}}, \label{19}\\
8\pi P_{\perp}=-\frac{1}{A^2}\left[\frac{\ddot R}{R}+\frac{{\ddot R}^{\prime}}{R^{\prime}}
-\frac{\dot A}{A}\frac{\dot R}{R}
-\left(\frac{\dot A}{A}
-\frac{\dot R}{R}\right)\frac{{\dot R}^{\prime}}{R^{\prime}}
\right] \nonumber\\
+\frac{1}{R^{\prime 2}}\left[\frac{A^{\prime\prime}}{A}-\left(\frac{R^{\prime\prime}}{R^{\prime}}
-\frac{R^{\prime}}{R}\right)\frac{A^{\prime}}{A}\right], \label{20}
\end{eqnarray}
while the mass function (\ref{17masa}) now reads,
\begin {equation}
m=\frac{R}{2}\left(\frac{\dot R}{A}\right)^2. \label{21}
\end{equation}

Using (\ref{18}) and (\ref{19}) with (\ref{17masa})  we obtain 

\begin{eqnarray}
D_Tm=-4\pi\left(P_rU+ qE\right)R^2.
\label{22Dt}
\end{eqnarray}

Three interesting features of Euclidean stars  deserve to be highlighted at this point.
\begin{itemize}
\item From (\ref{18}), it follows that if the system is dissipating   in the form of heat flow, the collapsing source requires $A^{\prime}\neq 0$, implying because of (\ref{5c}) $a^{\alpha}\neq 0$. This means that dissipation does not allow collapsing particles to follow geodesics. Inversely, of course, {\it non-dissipative Euclidean models are necessarily geodesic}, since $q=0$ implies because of (\ref{5c}) and   (\ref{18}) that $a^\alpha=0 $.

\item From  (\ref{21}) it follows that if ${\dot R}=0$ then $m=0$ and spacetime becomes Minkowskian. Therefore {\it all Euclidean stars are necessarily non-static}. Furthermore, using  (\ref{19n}),
 (\ref{21}) can be rewritten as
\begin{equation}
\frac{m}{R}=\frac{U^2}{2}. \label{24}
\end{equation}
The above equation  can be interpreted as the Newtonian kinetic energy (per unit mass) of the collapsing particles being equal to their Newtonian potential energy.

\item The proper radial three-acceleration $D_TU$ of an in falling particle inside $\Sigma$ can
be calculated to obtain
\begin{equation}
D_TU=-\left(\frac{m}{R^2}+4\pi P_r R\right)
+Ea. \label{28n}
\end{equation}

On the other hand, due  to the Euclidean condition, the dynamical equation (\ref{28n})  becomes
\begin{equation}
D_TU=-\left(\frac{m}{R^2}+4\pi P_r R\right)-\frac{8\pi qR}{2U}.
\label{3mnubis}
\end{equation}
Thus it appears that the contribution of  the non-gravitational force term (the last on the right hand side) to $D_TU$, for any fluid element, does not depend   on its inertial mass density ($\mu+P_r$).  Or, in other words,  non-gravitational forces produce a radial three-acceleration independent on  the inertial mass density of the fluid element, just as gravitational force term does.
\end{itemize}

\section{A hierarchy of Euclidean stars according to their complexity}

We shall now proceed to classify Euclidean stars by imposing different restrictions based on the notion of complexity discussed above.  Thus we shall start by considering the simplest possible model by assuming the vanishing complexity factor and the $H$ mode of evolution. From there on we shall relax the conditions on the complexity thereby   allowing more complex systems. We shall consider separately non--dissipative, and dissipative fluids.

Let us consider first the non-dissipative case.
\subsection{Evolution  with $q=0$}
As mentioned before, for this case we have from (\ref{18}) that $A^{\prime}=0$ which means $A=A(t)$ and by rescaling $t$ we can put without loss of generality
\begin{equation}
A=1.\label{25}
\end{equation}
Of course such models are members of  the Lema\^{\i}tre-Tolman-Bondi (LTB) spacetimes \cite{lemaitre,tolman,bondi}, furthermore they correspond to the parabolic case.

Indeed, the general metric  for LTB spacetimes read,
\begin{equation}
ds^2=-dt^2+\frac{R^{\prime 2}}{1-J(r)}dr^2+R^2(d\theta^2+\sin^2\theta d\phi^2),
\label{25IIIppnj}
\end{equation}
where $J(r)$ is an arbitrary function of $r$.

Imposing  the Euclidean condition (\ref{16}) in (\ref{25IIIppnj}), one obtains  $J=0$, which defines  parabolic LTB spacetimes. 

In this latter  case we know that for pure dust
   \begin{equation}
 R^{3/2}=(2m)^{1/2}\eta^3, \qquad \frac{2}{3}\eta^3=t-t_{bb}(r),
\label{int4}
\end{equation}
where $t_{bb}(r)$ is an integration function of $r$.

However the above solution is only valid for dust. For sake of completeness  we shall consider an anisotropic fluid (we recall that  LTB spacetime
is compatible with an anisotropic fluid \cite{sussman}.)
\subsubsection{Models satisfying $H$ condition and $Y_{TF}=0$.}
If we impose the $H$ condition then the shear vanishes and it follows from (\ref{itfm}) and (\ref{5b1}) that $Y_{TF}=0$. Furthermore since the fluid is geodesic, $QH$  and $H$  are equivalent. In such a case, as it has been shown in  \cite{7n}, the space--time is the FRW solution. 

Thus the simplest non-dissipative Euclidean star is described by the FRW line element.
\subsubsection{Models satisfying only $Y_{TF}=0$.}

In order to find the next subfamily of solutions in terms of increasing complexity,  let us consider that $Y_{TF}=0$  but the system does not evolve in the $H$ regime. Then from $Y_{TF}=0$  and (\ref{16}) it follows 
\begin{equation}
\ddot R=R \Psi(t),
\label{t1}
\end{equation}
where $\Psi(t)$ is an arbitrary function of its argument.

Next, from equations (\ref{19}), (\ref{20}) and (\ref{t1}) we find
\begin{equation}
8\pi\Pi=\frac{\dot R^\prime \dot R}{R R^\prime}-\left(\frac{\dot R}{R}\right)^2.
\label{t2}
\end{equation}

It is worth noticing that if we assume the fluid to be isotropic then it follows from (\ref{t2}) that $R$ is separable, i.e. $R=R_1(t) R_2(r)$ implying that the $QH$ ($H$)  condition is satisfied and we have FRW.

From the above it follows that we have to consider anisotropic models.

In the non--dissipative case ($q=0$) we have $m_\Sigma=M=$constant, then using (\ref{24}) evaluated on the boundary surface we may write
\begin{equation}
\dot R^2_\Sigma=\frac{2M}{R_\Sigma},
\label{ma2}
\end{equation}
whose solution reads
\begin{equation}
R_\Sigma=\left(\frac{M}{2}\right)^{1/3} \left[3(t-c_1)\right]^{2/3},
\label{mab}
\end{equation}
where $c_1$ is a constant of integration. This solution represents an ever expanding fluid configuration. Changing $t$ by $-t$  in (\ref{mab}) we obtain the corresponding collapsing solution.

Then using (\ref{mab}) in (\ref{t1}) evaluated on $\Sigma$, we obtain for $\Psi(t)$
\begin{equation}
\Psi(t)=-\frac{2}{9\left(t-c_1\right)^2} .
\label{ma3}
\end{equation}

From the results above and the field equations we may write
\begin{equation}
8\pi P_r=\frac{4}{9(t-c_1)^2}-\left(\frac{\dot R}{R}\right)^2,
\label{prrma}
\end{equation}

\begin{equation}
8\pi P_\bot= \frac{4}{9(t-c_1)^2}-\frac{\dot R^\prime}{R^\prime}\frac{\dot R}{R}\quad,
\label{ptma}
\end{equation}

\begin{equation}
8\pi \mu= 2\frac{\dot R^\prime}{R^\prime}\frac{\dot R}{R}+\left(\frac{\dot R}{R}\right)^2,
\label{roma}
\end{equation}

\begin{equation}
8\pi (P_r-P_\bot)= \frac{\dot R^\prime}{R^\prime}\frac{\dot R}{R}-\left(\frac{\dot R}{R}\right)^2 .
\label{roma}
\end{equation}

The general solution of (\ref{t1}), using  (\ref{ma3}) reads
\begin{equation}
R=(t-c_1)^{1/3}\left[ R_1(r)+R_2(r)(t-c_1)^{1/3}\right],
\label{nr1}
\end{equation}
where $R_1$ and $R_2$ are two arbitrary functions of integration.

It is a simple matter to check that (\ref{nr1}) solves (\ref{t1}) for the function $\Psi(t)$ given by (\ref{ma3}).

The mass function for this family of solutions reads
\begin{equation}
m=\frac{1}{18}\left[\frac{R_1+R_2(t-c_1)^{1/3}}{(t-c_1)^{1/3}}\right] \left[\frac{R_1+2R_2(t-c_1)^{1/3}}{(t-c_1)^{1/3}}\right]^2,
\label{mnr1}
\end{equation}
which asymptotically, as $t\rightarrow \infty$, becomes
\begin{equation}
m(\infty,r)=\frac{2}{9}R_2^3.
\label{mnr2}
\end{equation}

It is worth noticing that since we are considering non--vanishing pressure then in general $m$ also depends on $t$ as it follows from (\ref{22Dt}).

Also, since the system is non--dissipative then $m_\Sigma=M =$constant, implying
\begin{equation}
R_{1\Sigma}=0, \qquad R_{2\Sigma}=\left(\frac{9M}{2}\right)^{1/3}.
\label{bR}
\end{equation}

Thus the physical variables read

  \begin{equation}
    8\pi \mu=\frac{X}{9\left(t-c_1\right)^2}\left(X+2Y\right)=\frac{X}{9 r^2_\Sigma t^{\ast 2}}\left(X+2Y\right),\label{dennd1}
\end{equation}

\begin{equation}
    8\pi P_r=\frac{1}{9\left(t-c_1\right)^2}\left(4-X^2 \right)=\frac{1}{9 r^2_\Sigma t^{\ast 2}}\left(4-X^2\right),\label{prndn1}
\end{equation}

\begin{equation}
    8\pi P_\bot=\frac{1}{9\left(t-c_1\right)^2}\left(4-X Y\right)=\frac{1}{9 r^2_\Sigma t^{\ast 2}}\left(4-X Y\right),\label{ptndn1}
\end{equation}
  with
  \begin{equation}
X\equiv \frac{R_1+2 R_2(t-c_1)^{1/3}}{R_1+R_2(t-c_1)^{1/3}},\quad Y\equiv \frac{R^\prime_1+2 R^\prime_2(t-c_1)^{1/3}}{R^\prime_1+ R^\prime_2(t-c_1)^{1/3}},
\label{in}
\end{equation}
and  $t^\ast\equiv (t-c_1)/r_\Sigma$.

From (\ref{bR}) it follows that $X_\Sigma=2$ implying $P_{r\Sigma}=0$ as expected from (\ref{j3}). 

In order to specify a model, let us propose

\begin{equation}
     R_1(r)=\left(r^2-r r_\Sigma\right)^{1/3} \equiv r^{2/3}_\Sigma \left(z^2-z\right)^{1/3} ,
     \label{r1m}
     \end{equation}
    \begin{eqnarray}
     R_2(r)=\left(r^2-r r_\Sigma+\frac{81M^2}{4}\frac{r}{r_\Sigma}\right)^{1/6}\nonumber \\ \equiv r^{1/3}_\Sigma \left(z^2-z+\frac{81}{4} M^{\ast 2} z\right)^{1/6} \label{r2m},
\end{eqnarray}
  which satisfies the boundary condition (\ref{bR}), and where $z\equiv r/r_\Sigma$, and $M^\ast \equiv M/r_\Sigma$.
  
  In this model the variables $X, Y$ become
  \begin{equation}
  X=\frac{\left(z^2-z \right)^{1/3}+2\left(z^2-z+\frac{81 M^{\ast 2} z}{4}\right)^{1/6} t^{\ast1/3}}{\left(z^2-z \right)^{1/3}+\left(z^2-z+\frac{81  M^{\ast 2} z}{4}\right)^{1/6} t^{\ast 1/3}},
  \label{xm}
  \end{equation}

  \begin{equation}
  Y=\frac{W+2 V}{W+V},
  \label{ym}
  \end{equation}

where
\begin{equation}
W\equiv 2\left(2z-1\right)\left(z^2-z+\frac{81}{4}M^{\ast 2}z\right)^{5/6},
\label{w}
\end{equation}
and
\begin{equation}
V\equiv \left(2z-1+\frac{81 M^{\ast 2} }{4}\right)\left(z^2-z \right)^{2/3} t^{\ast 1/3}.
\label{V}
\end{equation}

Finally, for the mass function we obtain the expression
\begin{equation}
m^\ast=\frac{18m}{r_\Sigma}=\frac{\left(Q S^2\right)}{t^\ast},
\label{mmr}
\end{equation}
where
\begin{equation}
Q\equiv \left(z^2-z\right)^{1/3}+t^{ \ast 1/3}\left(z^2-z+\frac{81}{4} M^{\ast 2} z\right)^{1/6},
\label{Q}
\end{equation}
and
\begin{equation}
S\equiv \left(z^2-z\right)^{1/3}+2t^{ \ast 1/3}\left(z^2-z+\frac{81}{4}M^{\ast 2} z\right)^{1/6}.
\label{S}
\end{equation}

\begin{figure}[h]
\includegraphics[scale=0.7]{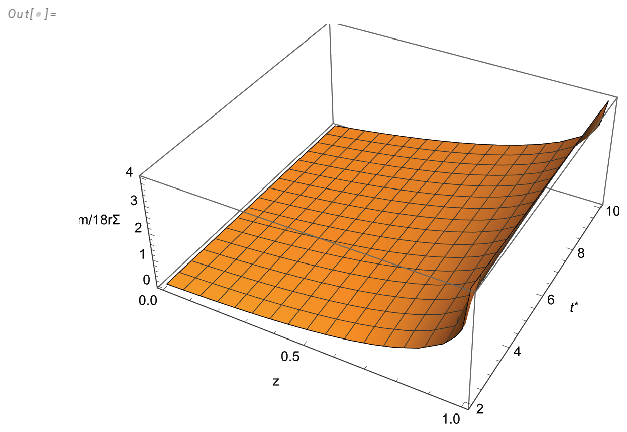}
\caption {\it  $m^\ast$ as function of $z$ in the interval $[0,1]$, and
 $t^\ast$ in the interval $[2,10]$}\label{fig 2}
\end{figure}

\begin{figure}[h]
\includegraphics[scale=0.7]{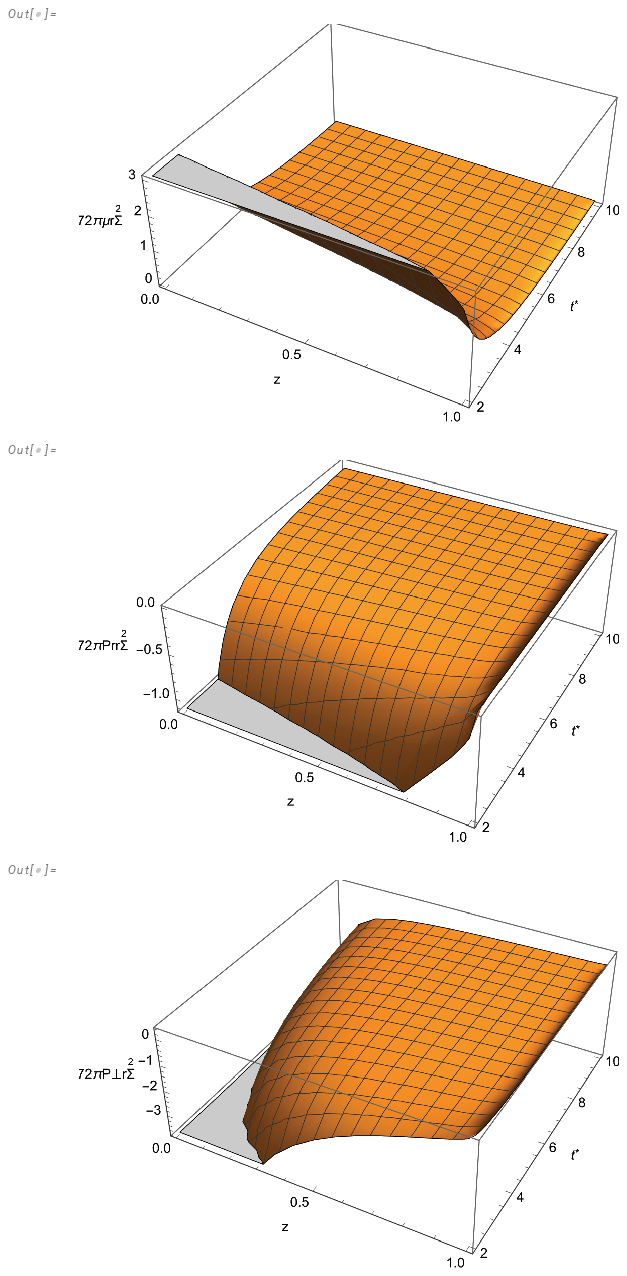}
\caption {$72\pi \mu  r^2_{\Sigma^{(e)}}$, $72\pi P_r  r^2_{\Sigma^{(e)}}$ and $72\pi P_\bot  r^2_{\Sigma^{(e)}}$  as functions of $z$ in the interval $[0,1]$, and  $t^\ast$ in the interval $[2,10]$.}\label{fig 1}
\end{figure}

Figure \ref{fig 2} and Figure \ref{fig 1}  depict the evolution and spatial distribution of the mass function and the physical variables for this model.

\subsection{Evolution with $q\neq 0$}
We shall next consider the dissipative case. 

Let us starting by noting that  the $H$ condition (as shown in \cite{7n}) implies that the fluid is geodesic, which is incompatible with a dissipative Euclidean star. Then,  concerning complexity, the lowest degree of complexity admissible for a dissipative Euclidean star would be described by $QH$ condition and $Y_{TF}=0$.

We shall consider here two possible families of solutions with different degree of complexity. We shall first consider  fluid distributions satisfying  the vanishing complexity factor condition plus the $QH$ condition. Such a case, as mentioned above represents models with the lowest  degree of  complexity.  Next,  we shall consider
the family of solutions satisfying  the vanishing complexity factor condition alone. In order to integrate the equations we have complemented this latter condition  with  the shear-free condition. 

\subsubsection{Models with $Y_{TF}=0$ and $QH$ evolution.}

As it follows from  (\ref{ch1}), the ($QH$) condition reads 
\begin{equation}
\dot R=A \omega R
\label{dis2s}
\end{equation}

where $\omega$ is function of integration, with dimensions $1/[length]$, which without lost of generality may be put equal to $1$ by a reparametrization of  $t$. 

Let us now impose the condition $Y_{TF}=0$, which using (\ref{itfm}) and (\ref{dis2s}) reads
\begin{eqnarray}
\frac{\omega^2 R^2}{\dot R^2}\left[\frac{\ddot R}{R}-\frac{ \ddot R^\prime}{R^\prime}+\left(\frac{\ddot R}{\dot R}-\frac{\dot R}{R}\right)\left(\frac{\dot R^\prime}{R^\prime}-\frac{\dot R}{R}\right) \right]\nonumber \\ +\frac{R}{\dot R (R^\prime)^2}\left[\left(\frac{\dot R}{R}\right)^{\prime \prime} -\left(\frac{\dot R}{R}\right)^\prime \left(\frac{R^{\prime \prime}}{R^\prime}+\frac{R^\prime}{R}\right)\right]=0.
\label{dis1bb}
\end{eqnarray}

In order to find a solution to the above equation, we shall split it into a system of two equations resulting from equating to zero the expressions within each one of  square brackets in (\ref{dis1bb}), producing  
\begin{equation}
\dot H^\prime=H^\prime\left(\frac{\dot H}{H}-2 H\right)=0,
\label{idsp1}
\end{equation}
and 
\begin{equation}
\frac{H^{\prime \prime}}{H^\prime}=\frac{K^\prime}{K}+2 K,
\label{idsp2}
\end{equation}
where $H\equiv \frac{\dot R}{R}$ and $K\equiv\frac{R^\prime}{R}$.

Equation (\ref{idsp2}) may be solved to obtain
\begin{equation}
R^2=\frac{F_1(t)}{F_2(t)+G(r)},
\label{dsp3}
\end{equation}
where $F_1(t), F_2(t), G(r)$ are arbitrary functions of their argument. $F_2$ and $G$ are dimensionless, whereas $F_1$ has dimensions $[length^2]$.

Thus, so far our model is defined by (\ref{16}), (\ref{dis2s}), and (\ref{dsp3}).

Let us now integrate (\ref{idsp1}), whose first integral reads 

\begin{equation}
    \frac{H^\prime}{H}=\frac{g_1(r)}{R^2}\label{psol},
\end{equation}

\noindent where  $g_1(r)$ is an arbitrary function of $r$. 

Taking the $r$ derivative of (\ref{psol}) and using (\ref{idsp2}) we obtain 

\begin{equation}
    \frac{K^\prime}{K}-\frac{H^\prime}{H}+4K-\frac{g_1^\prime}{g_1}=0,\label{zy}
\end{equation}
whose solution may be written as

\begin{equation}
    K=\frac{g_1(r) g_2(t)H}{R^4}\, ,\label{sz}
\end{equation}

\noindent where  $g_2(t)$  is an arbitrary function of $t$.

\noindent Using   (\ref{sz}) in  (\ref{zy}) produces

\begin{equation}
    (R^2)^\prime R^4-g_1(r)g_2(t) (R^2 \dot )=0,\label{RR}
\end{equation}
or, introducing the variables

\begin{equation}
    \epsilon=R^2\,,\quad \tau=\int \frac{dt}{g_2(t)}\, , \quad \tilde r=\int g_1(r)dr\, , \label{cambios}
\end{equation}
it becomes
\begin{equation}
    \frac{\partial \epsilon}{\partial \tau}-\epsilon^2 \frac{\partial \epsilon}{\partial \tilde r}=0, 
\end{equation}
whose solution reads
\begin{equation}
    \tilde r=-\tau \epsilon^2+\Phi (\epsilon),
\end{equation}

\noindent where  $\Phi$ is an arbitrary  function. 

Going back to the original variables the above equation reads

\begin{equation}
    \int g_1(r)dr=-R^4\int \frac{dt}{g_2(t)} +\Phi(R^2).
\end{equation}

\noindent Each solution corresponding to a given choice of $\Phi$.

Next, we may find a relationship between the time dependent functions appearing in (\ref{dsp3}), by imposing the boundary condition (\ref{j3}), which now reads
\begin{equation}
\frac{4 \dot F}{F}=\frac{-3 \omega^2 (F_1/F)^{.}}{1+\omega\sqrt{F_1/F}},
\label{b1m}
\end{equation}
whose integration produces
\begin{equation}
4\ln F=-6\omega\sqrt{F_1/F}+6\ln{\left(\omega+\omega^2\sqrt{F_1/F}\right)}+ \xi,
\label{b2m}
\end{equation}
where $F\equiv F_2+G(r_\Sigma)$ and $\xi$ is a constant of integration.

Thus the whole family of solutions satisfying  the $Y_{TF}=0$ and $QH$ evolution condition (besides de Euclidean condition), depends on one arbitrary function of $t$ and one arbitrary function on $r$. 

In terms of the function $R$ the physical variables for this family read
\begin{equation}
8\pi\mu =\omega^2\left(1+2\frac{{\dot R}^{\prime}}{R^{\prime}} \frac{R}{\dot R}\right), \label{17mo}
\end{equation}
\begin{equation}
8\pi q=2\omega \left(\frac{-\dot R^\prime}{R^\prime \dot R}+\frac{1}{R}\right), \label{18mo}
\end{equation}
\begin{equation}
8\pi P_r=-3\omega^2+\frac{2\dot R^\prime}{R\dot R R^\prime}-\frac{2}{R^2} ,\label{19mo}
\end{equation}
\begin{equation}
8\pi P_\bot=-\omega^2\left(\frac{\ddot R^\prime R^2}{R^\prime \dot R^2}+1-\frac{\ddot R \dot R^\prime R^2}{\dot R^3 R^\prime}+\frac{2\dot R^\prime R}{R^\prime \dot R}\right)+\frac{2\dot R^\prime}{R \dot R R^\prime}-\frac{2}{R^2}\label{20mo}.
\end{equation}

In order  to illustrate  how to find a specific model of this family, we shall present in the  next subsection a toy model.

\subsubsection{A toy model}
We start the construction of this model by assuming
\begin{equation}
\frac{F_1}{F}=b t,
\label{toy1}
\end{equation}
where $b$ is a constant with dimensions of $[length]$.

Then using (\ref{toy1})  in (\ref{b2m}) we obtain
\begin{equation}
F=e^{-(3/2) \omega \sqrt{b t}}[1+w\sqrt{b t}]^{3/2},
\label{toy2}
\end{equation}
where for convenience we have chosen $-6\ln w=\xi$.

In order to fully specify the model, it remains to provide the function $G(r)$. For simplicity, and to avoid shell crossing singularities we shall assume
\begin{equation}
G=-\alpha r
\label{toy3}
\end{equation}
where $\alpha$ is a positive constant  with dimensions $[1/length]$. 

Using the above expressions, the areal radius of the boundary surface becomes
\begin{equation}
\omega R_\Sigma =t^{\ast 1/2}
\label{toyR}
\end{equation}
whereas for the physical variables we obtain
\begin{equation}
\frac{8\pi \mu}{\omega^2}=\frac{3\left\{(4 \Omega-3 t^{\ast})[\Omega^{3/2}+(1-z) e^{(3/2)t^{\ast 1/2}}]+7t^{\ast} \Omega^{3/2} \right\}}{(4 \Omega-3 t^{\ast})[\Omega^{3/2}+(1-z) e^{(3/2)t^{\ast 1/2}}]+3t^{\ast} \Omega^{3/2} },
\label{toy4}
\end{equation}

\begin{equation}
\frac{8\pi q}{\omega^2}=\frac{-12t^{\ast 1/2} \Omega^{3/4}\left[\Omega^{3/2}+(1-z)e^{(3/2)t^{\ast 1/2}}\right]^{1/2}}{(4 \Omega-3 t^{\ast})\left[\Omega^{3/2}+(1-z)e^{(3/2)t^{\ast 1/2}}\right]+3t^{\ast} \Omega^{3/2}},
\label{toy5}
\end{equation}

\begin{equation}
\frac{8\pi P_r}{\omega^2}=-3+\frac{12 \left[\Omega^{3/2}+(1-z)e^{(3/2)t^{\ast 1/2}} \right] }{(4 \Omega-3 t^{\ast})\left[\Omega^{3/2}+(1-z)e^{(3/2)t^{\ast 1/2}}\right]+3t^{\ast} \Omega^{3/2}},
\label{toy6}
\end{equation}
where  $\Omega\equiv 1+t^{\ast 1/2}$;  $t^\ast \equiv  \omega^2 b t$; $z\equiv \frac{r}{r_\Sigma}$, and we assume $\alpha r_\Sigma=1$.

The temperature and the tangential pressure for this model may be calculated using (\ref{V2}), (\ref{toy5}) and (\ref{20mo}), however the resulting expressions are cumbersome and not very illuminating. Accordingly we will dispense  with these  expressions and the corresponding graphics. 

\begin{figure}[h]
\includegraphics[scale=0.7]{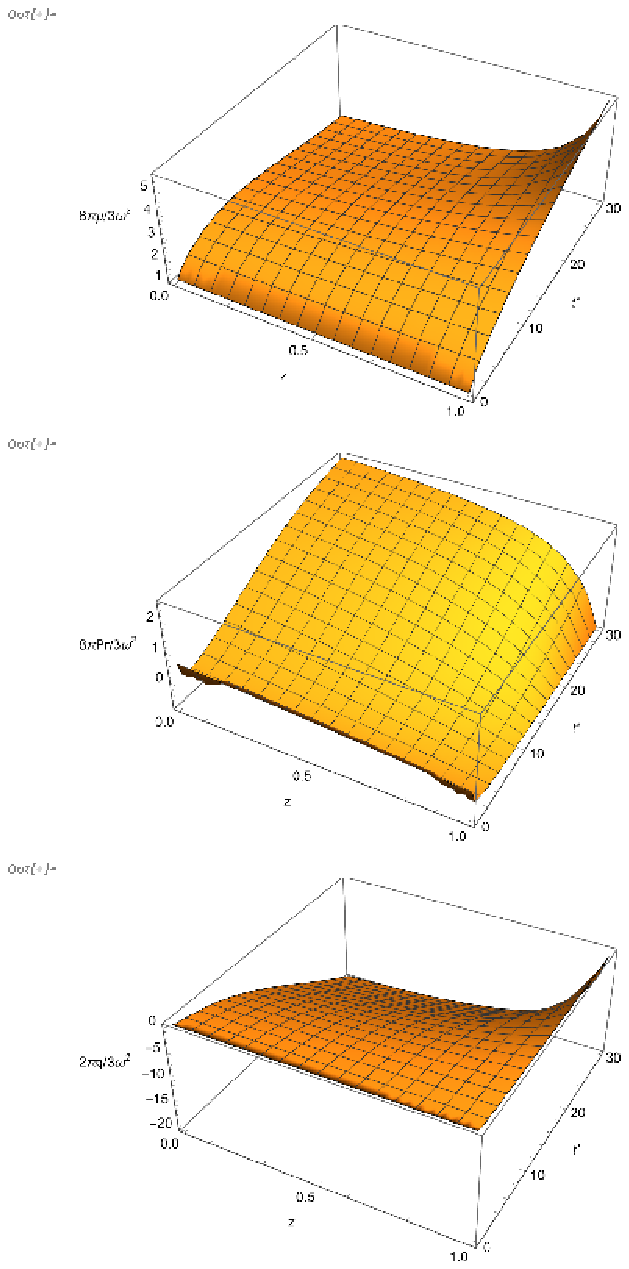}
\caption {$8\pi \mu/\omega^2$, $8\pi P_r /3\omega^2$ and $2\pi q/3\omega^2$  as functions of $z$ in the interval $[0,1]$, and  $t^\ast$ in the interval $[0,30]$.}\label{fig 3}
\end{figure}

Figure \ref{fig 3}  depicts the evolution and spatial distribution of $\mu, P_r,  q$ for this model, which represents an ever expanding sphere starting from a singularity.

\subsubsection{Models with $Y_{TF}=\sigma= 0$.}
We shall now relax the condition of $QH$ evolution. In  order to be able to integrate the system of field  equations we need an additional condition, which we choose to be the shear--free condition. The motivation of assuming such a restriction stems from the role played by shear in the collapse of dissipative fluid spheres  (see \cite{sh1,shen,sh2} and references therein).
Obviously in this case we cannot assume $QH$ since that would imply, due to (\ref{ch1}) that $q=0$. 

The general shear-free case for dissipative Euclidean stars has been considered in \cite{pe}.  In such a  case the line element can be written as  \cite{Glass}
\begin{equation}
ds^2-=-A^2dt^2+B^2[dr^2+r^2(d\theta^2+\sin^2\theta d\phi^2)], \label{35}
\end{equation}
and the Euclidean condition becomes
\begin{equation}
B=f(t),
\label{eu}
\end{equation}
implying
\begin{equation}
R=f(t)r, \label{36}
\end{equation}
where $f$ is an arbitrary function of $t$.

The field equations (\ref{17}-\ref{20}) now read,
\begin{eqnarray}
8\pi\mu &=&\frac{3}{A^2}\left(\frac{\dot f}{f}\right)^2, \label{37}\\
8\pi q &=&-2\frac{\dot f}{f^2}\frac{A^{\prime}}{A^2}, \label{38}\\
8\pi P_r&=&-\frac{1}{A^2}\left[2\frac{\ddot f}{f}-\left(2\frac{\dot A}{A}
-\frac{\dot f}{f}\right)\frac{\dot f}{f}\right]\nonumber \\&+&\frac{2}{f^2r}\frac{A^{\prime}}{A}, \label{39}\\
8\pi P_{\perp}&=&-\frac{1}{A^2}\left[2\frac{\ddot f}{f}-\left(2\frac{\dot A}{A}
-\frac{\dot f}{f}\right)\frac{\dot f}{f}\right]\nonumber \\&+&\frac{1}{f^2}\left(\frac{A^{\prime\prime}}{A}
+\frac{1}{r}\frac{A^{\prime}}{A}\right). \label{40}
\end{eqnarray}
From (\ref{39}) and (\ref{40}) we have
\begin{equation}
8\pi(P_{\perp}-P_r)=\frac{1}{f^2A}\left(A^{\prime\prime}-\frac{A^{\prime}}{r}\right). \label{41}
\end{equation}

On the other hand, using (\ref{eu}) and (\ref{36}) in (\ref{itfm}), the condition $Y_{TF}=0$ becomes
\begin{equation}
A^{\prime \prime}=\frac{A^\prime}{r},
\label{dis4s}
\end{equation}
whose solution reads
\begin{equation}
A=\gamma(t) r^2+\delta(t),
\label{dis5s}
\end{equation}
where $\gamma$ and $\delta$ are two functions of integration

From (\ref{41}), (\ref{dis4s}) it follows that the fluid is isotropic, and  from (\ref{EY}) it follows that it is also conformally flat.

The general form of all conformally flat and shear-free metrics is known \cite{hlsw}, it reads
\begin{equation}
A=\left[e_1\left(t\right)r^2+1\right]B, \label{II2}
\end{equation}
where $e_1$ is an arbitrary function of $t$,
and
\begin{equation}
B=\frac{1}{e_2(t)r^2+e_3(t)}, \label{II4}
\end{equation}
where $e_2$ and $e_3$ are arbitrary functions of $t$.

The Euclidean condition and the vanishing complexity factor condition then imply
\begin{equation}
e_2=0, \;\; e_3=\frac{1}{f},\;\; e_1=\frac{\gamma}{f},\;\; \delta=f.
\label{met}
\end{equation}

 An approximate solution of this kind has been presented and discussed in \cite{hlsw}. Furthermore  an exact solution is also known \cite{GM}, which in turn is a particular  case of a family of solutions found in \cite{hja}. It reads (see Case III in \cite{GM})
\begin{equation}
f(t)=(\beta_1+\beta_2)^2 e^{-2\alpha r_{\Sigma} t},
\label{f}
\end{equation}
and
\begin{equation}
A=(\alpha r^2+1)f, \label{42}
\end{equation}
where $\alpha$, $\beta_1$ and  $\beta_2$ are constants.
The above solution satisfies junction conditions and its physical properties have been discussed in \cite{GM}. Thus we shall not  elaborate any further on it. Let us just mention that  its physical properties are reasonable and using (\ref{V2}) it has been shown that  relaxation effects plays an important role in the evolution of the system.

\section{Discussion and Conclusions}
The family of Euclidean stars encompasses a large number of solutions to Einstein equations describing the evolution of spherically symmetric fluid distributions, which may be dissipative or not. Research  work devoted to the application of some of these solutions in the study of the structure and evolution of compact objects may be found in \cite{pe,5e,9e,6e,7e,8e,4e,2e,3e,1e}  and references therein.

The main interest of these solutions stems from the fact that the heuristic condition defining these models is borrowed from Euclidean geometry, thereby allowing for the analytical modeling of stellar evolution  relying on a condition which  should hold at least for non very strong fields. Besides, it is worth noticing that such relevant solutions as FRW and some  LTB  models, are members of this family.

Our goal in this  work has been to introduce a hierarchical structure of different sub-families of Euclidean stars, based on their degree of complexity.  As a  measure of complexity we have resorted to notion of the complexity factor defined in \cite{6n,7n}, and to the notion of Homologous and Quasi-homologous evolution as defined in \cite{7n,7nn}.

We have considered, both the adiabatic and the dissipative case. In the former case it appears that the fluid must be geodesic, and therefore it belongs to the family of LTB solutions. More specifically to the parabolic case of LTB models.  
 
The highest degree of simplicity in this case, which correspond to the vanishing complexity factor and $H$ evolution, leads to the FRW space--time. We have next relaxed the $H$ condition, maintaining the vanishing of the complexity  factor condition. In such a case we found a family of solutions described by equations (\ref{dennd1})--(\ref{in}), depending on two arbitrary functions of $r$.

A specific model of the above mentioned family of solutions was obtained from the additional assumptions given by equations  (\ref{r1m}), (\ref{r2m}). Figure \ref{fig 2} and Figure \ref{fig 1} depict the evolution and spatial distribution of the mass function and the physical variables for this model. It is worth stressing that our motivation to present such a model was not to describe  any specific astrophysical or cosmological  scenario, but just to illustrate the way of obtaining  models.   The possible application of any  model  belonging to this family of solutions is out of the scope of this manuscript.

Next, we considered the dissipative case. The first family of solutions of this case satisfies the conditions $Y_{TF}=0$ and $QH$. This is the simplest family of Euclidean dissipative stars. 
In general this family of solutions depends on one arbitrary function of $t$ and one arbitrary function of $r$. A toy model belonging to this family was constructed, by assuming for the two functions mentioned above, the conditions (\ref{toy1}) and (\ref{toy3}). The behavior of the physical variables for this model is depicted in Figure \ref{fig 3}. Once again, we should stress the fact that this toy model was obtained just to illustrate the way of obtaining specific solutions corresponding to this family.

Finally, as the last step in  order  of increasing complexity we relaxed the $QH$ condition and considered only the $Y_{TF}=0$ condition. To integrate the system we complemented  this latter condition with the shear--free condition, which by the way  prevents the system to satisfy the $QH$ condition. Thus from the point  of view of complexity this family of solutions satisfy only one of the restrictions based on the notion of complexity. However, as shown above, this family of solutions is also conformally flat, and as a matter of fact was already known from previous works, and their properties have been analyzed in detail in   \cite{hlsw,GM,hja}.

We would like to conclude by emphasizing the main goal of our endeavor.  We looked for a   hierarchy of Euclidean stars  according to their complexity, as a byproduct of this research   we were able to present an analytical  method for the modeling of  these stars. The presented examples were proposed just to illustrate the method. In order to obtain models which could be applied to study specific astrophysical or cosmological scenarios it would be  necessary to relate the arbitrary functions involved in the different families of solutions to relevant observables variables, among which the gravitational surface redshift and the rate of emitted energy are probably the best candidates (although not the only ones).

\end{document}